\documentclass[aps,10pt,pre,twocolumn,amsmath,amssymb,amsfonts]{revtex4-2}

\usepackage{amssymb}
\usepackage{amsmath}
\usepackage{graphicx}
\usepackage{hyperref}

\def\Eq#1{Eq.~(\ref{#1})}
\def\Eqs#1{Eqs.~(\ref{#1})}
\def\Fig#1{Fig.~\ref{#1}}	
\def\Figs#1{Figs.~\ref{#1}}	

\def\no{\nonumber\\}

\def\>{\rangle}
\def\<{\langle}

\def\g{\gamma}
\def\k{\kappa}
\def\s{\sigma}
\def\m{\mu}
\def\lam{\lambda}
\def\t{\tau}
\def\Del{\Delta}
\def\bD{\bar{\Del}}
\def\al{\alpha}
\def\d{\partial}
\def\T{\text{T}}
\def\u{\boldsymbol{u}}
\def\v{\boldsymbol{v}}
\def\q{\boldsymbol{q}}
\def\L{\boldsymbol{L}}
\def\M{\boldsymbol{M}}
\def\tr{\text{tr}}

\begin{document}

\title{Entanglement generation across exceptional points in two-qubit open quantum system -- the role of initial states}

\author{B.~A.~Tay}
\email{BuangAnn.Tay@nottingham.edu.my}
\affiliation{School of Mathematical Sciences, Faculty of Science and Engineering, University of Nottingham Malaysia, Jalan Broga, 43500 Semenyih, Selangor, Malaysia}
\author{Yee Shean H'ng}
\affiliation{Department of Mechanical, Materials and Manufacturing Engineering, Faculty of Science and Engineering, University of Nottingham Malaysia, Jalan Broga, 43500 Semenyih, Selangor, Malaysia}

\date{\today}

\begin{abstract}
We study an open quantum system of two qubits that are coupled by swapping interaction.
Using the coupling strength between the qubits as a time scale, the Liouvillian of the system has exceptional points that depend on the disparity between the decay rates of the qubits. 
We find that the configuration of the initial states plays an important role in deciding the character of the entanglement dynamics at the initial stage of evolution.
Depending on whether or not the initial excitations of the qubits can be swapped by the interaction that couples them, a change in the total decay rate can be either consistently unfavorable to entanglement generation, or shift the dynamics from hindering to enhancing entanglement generation, or vice versa, as the system traverses the exceptional points.
The shift could also occur in a wide range of mixed states. We clarify the origin of the behavior in this work.

\end{abstract}

%\pacs{05.70.Ln,03.67.Mn}
%\keywords{Quantum entanglement;Open quantum system;}

\maketitle

%%%%%%%%%%%%%%%%%%%%%%%%%%
%       Section          %
%%%%%%%%%%%%%%%%%%%%%%%%%%
\section{Introduction}

Recently, non-Hermitian quantum systems \cite{Bender98,Bender02,Moiseyev11} have been effectively realized in the laboratories, for example, in three-level superconducting transmon circuits \cite{Naghiloo19,Chen21}, two-level atoms in cavity \cite{Han23}, photonic systems \cite{Ozdemir19,ZhangXinchen24}, optical systems \cite{Ruter10,Miri19} and mechanical systems \cite{Bender13,Zhou23}. The development provides avenue to study non-Hermitian degeneracy \cite{Berry04,Heiss12} that occurs at the exceptional points in the parameter space of the systems. At the exceptional points, the eigenvalues and the eigenvectors of the non-Hermitian Hamiltonian coalesce. In contrast to Hermitian degeneracy where the eigenvectors can still be chosen to be orthogonal, non-Hermitian degeneracy leads to a Jordan block structure of the Hamiltonian, giving rise to generalized eigenvectors \cite{Bhamathi96,Bohm97,Weintraub08}.

Initial studies on non-Hermitian degeneracy focused on the Hamiltonian level \cite{Bender98,Moiseyev11}. Later on, Liouvillian exceptional points were investigated \cite{Hatano19,Nori19,Chen22}. On the Liouvillian level, the quantum jump formalism offers insight in clarifying the differences between the exceptional points structure in non-Hermitian Hamiltonian and its related Liouvillian with jump terms \cite{Nori20b}.

Contrary to the common belief that dissipation is unfavorable to quantum processes, interesting phenomena were reported when the systems were close to the vicinity of non-Hermitian exceptional points, such as speeding up of entanglement generation \cite{Li23,ZhangJi24}, topological energy transfer \cite{Xu16,Assawaworrarit17}, enhanced sensing \cite{Hodaei17,Chen17}, or when the dynamics encircles the exceptional points \cite{Berry11}, leading to chiral state transfers \cite{Uzdin11,Sun23} and enhancement of quantum heat engine \cite{Zhang22}.

A recent study of the effective non-Hermitian Hamiltonian of two qubits derived from three-level system subjected to resonant drives, revealed that at the vicinity of the exceptional point which is decided by the driving amplitude \cite{Li23}, a weak coupling between the qubits could greatly enhance entanglement generation in a very short time scale. The general entanglement dynamics of this system was studied later \cite{Zhang24}. Different types of entanglement dynamics can be induced by exceptional points through adjustment of the driving amplitude \cite{Li25}.

Motivated by this result, we investigate a much simpler system of two qubits coupled by swapping interaction, and explore the role of initial states on entanglement generation while the system traverses exceptional points.
We find that the initial states, depending on whether or not their initial excitations can be swapped by the interaction that couples the two qubits, play an essential role in deciding the extent of entanglement generation at the initial stage of the evolution. 
For initial conditions with one of the qubits excited, the excitation can be swapped between the two qubits.
Then an increase in decay rate consistently hinders entanglement generation as the system crosses over the exceptional points. 
On the other hand, when both qubits are initially excited, the swapping of excitations does not bring observable changes to the state. 
Now an increase in the decay rate surprisingly enhances entanglement generation at the initial stage of the dynamics.
However, when the system crosses over the exceptional point, the dynamics turns over into impeding entanglement generation. This is contrary to the common belief that an increase in decay rate is in general unfavorable to entanglement generation.

Though arising from different origins, the effect reminds us of the noise-assisted energy transport \cite{Plenio08,Plenio10,Plenio12} discussed recently in quantum networks and light-harvesting complexes, where dissipation can occasionally be advantageous to energy transport.
Here, our focus is on entanglement generation.
We clarify the origin of the effect in subsequent sections.

%{\color{blue}
The model of two qubits coupled via the swapping interaction as well as its extension to non-Hermitian qubits, a chain of qubits and their variants has been used widely to model various quantum processes, such as the dynamics of quantum entanglement \cite{Han23,Li23,Zhang24,Li25}, energy transport in quantum networks and photosynthetic systems \cite{Plenio08,Plenio10,Plenio12}, efficiency of quantum heat engines \cite{Hewgill18,Bresque21}, quantum sensing in photonic systems \cite{Miri19,Wiersig20,Li23}, exciton transfer in molecular systems \cite{May11}, etc.
Here, our focus is on the behavior of entanglement generation between two coupled qubits under different initial conditions.
Since entanglement is an important resource in quantum information processing \cite{Nielsen}, our results provide a different perspective on the interplay between dissipation and entanglement generation, where dissipation could be beneficial to entanglement generation under some initial states.
%}

In Sec.~\ref{SecH}, we introduce the quantum master equation of two qubits and clarify the structure of its exceptional points in the subspace spanned by the $X$ state. Then we elucidate the behaviors of the dynamics with either one or two qubits excited as the initial conditions in Sec.~\ref{SecIni10} and \ref{SecIni11}, respectively. We show that the time evolution under different initial conditions exhibits different behaviors when the decay rate changes across the exceptional points. 
A mixture of different initial conditions is considered in Sec.~\ref{SecIniMixed}.
We then summarize our results in Sec.~\ref{SecConc}.

%%%%%%%%%%%%%%%%%%%%%%%%%%
%       Section          %
%%%%%%%%%%%%%%%%%%%%%%%%%%
\section{Two-qubit open quantum system}
\label{SecH}

%{\color{blue}
The quantum master equation we look into is a Gorini-Kossakowski-Sudarshan-Lindblad (GKSL) equation with completely positive evolution \cite{Kossa76,Lindblad76}. Variants of the systems had been previously used to study entanglement sudden death in various setups \cite{Yu04,Yu10}. 
However, the behavior of entanglement generation across exceptional points has not been fully explored.
Previous studies \cite{Li23,Zhang24,Li25} considered a model with resonant drives and qubits with equal decay rates. Here, we exclude resonant drives and permit unequal decay rates of the qubits to study the effects on entanglement generation.
%}

We consider two identical qubits with effective energy $\omega_0$, using the units $\hbar=1$. They are coupled by a swapping interaction with strength $J$ \cite{Li23},
%%%
\begin{align}   \label{H}
    H&=\frac{\omega_0}{2}\s_z +\frac{\omega_0}{2}\m_z
        +J\big(\s_+ \m_- +\s_- \m_+ \big)\,,
\end{align}
%%%
where $\s_z$ and $\m_z$ label the $z$-component of the Pauli matrices for the first and second qubit, respectively, and $\s_\pm\equiv(\s_x\pm i\s_y)/2$ are the creation and annihilation operators of the first qubit, with similar definitions $\m_\pm$ for the second qubit.
The interaction can be realized between two identical two-level atoms interacting with a cavity mode \cite{Zheng00}, or between two non-Hermitian qubits in a superconducting circuit \cite{Han23}.
We consider two qubits free from resonant drives. The first and second excited qubits relax to their ground states with decay rates $\g_1$ and $\g_2$, respectively. In the interaction picture, the two-qubit reduced dynamics is described by the GKSL master equation
%%%
\begin{align}   \label{rhot}
    \frac{\d\rho}{\d t}&=-K\rho\,,
\end{align}
%%%
where
%%%
\begin{align}   \label{K}
    K\rho&=i[J\big(\s_+ \m_- +\s_- \m_+),\rho]\no
    &-\frac{\g_1}{2}\big(2\s_-\rho\s_+ -\s_+\s_-\rho-\rho\s_+\s_-\big)\no
    &-\frac{\g_2}{2}\big(2\m_-\rho\m_+ -\m_+\m_-\rho-\rho\m_+\m_-\big)
    \,.
\end{align}
%%%

The two-qubit density operator, $\rho$, is a $4\times4$ positive semidefinite Hermitian matrix with trace 1. In the following discussion, we will restrict our attention to the subspace spanned by the so-called $X$ state \cite{Yu04},
%%%
\begin{align}   \label{Xst}
    \rho_X=\left(\begin{array}{cccc}
            a &0 &0 &h\\
            0 &b &m &0 \\
            0 &m^* &c &0 \\
            h^* & 0 &0 &d
          \end{array} \right)\,,
\end{align}
%%%
with the normalization condition $a+b+c+d=1$, and $*$ denotes complex conjugate.
The $X$ state remains in this subspace under the evolution of the reduced dynamics \eqref{rhot}. 
Introducing a dimensionless time
%%%
\begin{align}   \label{tau}
    \t&\equiv 2Jt\,,
\end{align}
%%% 
and defining three parameters,
%%%
\begin{align}   \label{x}
    x&\equiv b+c\,,
\end{align}
%%%
%%%
\begin{align}  \label{y}
    y&\equiv b-c\,,
\end{align}
%%%
%%%
\begin{align}   \label{z}
    z&\equiv m-m^*\,,
\end{align}
%%%
the equation of motion of the matrix elements together with their solutions is listed in App.~\ref{AppSol}.
We further introduce a rescaled total decay rate
%%%
\begin{align}   \label{gam}
        \g&\equiv\frac{1}{4J}(\g_1+\g_2)\,,
\end{align}
%%%
and a rescaled disparity between the two qubits' decay rates
%%%
\begin{align}   \label{kap}
    \k&\equiv\frac{1}{4J}(\g_2-\g_1)\,.
\end{align}
%%%
Because $\g_1,\g_2\geq 0$, the following constraint must be satisfied by the rescaled quantities,
%%%
\begin{align}   \label{g<k}
        |\k|\leq \g\,.
\end{align}
%%%
Two of the parameters, $m+m^*$ and $h$ (and its complex conjugate), evolve independently from each other, see \Eqs{at}-\eqref{ht} for the details.
Hence, we focus on the subspace formed by the five parameters $a,x,y,z$ and $d$.

The generator of evolution for the matrix elements of this subspace has the eigenvalues,
%%%
\begin{align}   \label{eigenv}
    \lam_0=0,\quad \lam_1=-2\g , \quad \lam_2=-\g,\nonumber\\ 
    \lam_3=-\g- \Del, \quad \lam_4=-\g+\Del\,,
\end{align}
%%%
see App.\ref{AppEigev}, where
%%%
\begin{align}   \label{Del}
    \Del&\equiv\sqrt{\k^2-1}
\end{align}
%%%
determines the behavior of the dynamics. 
When $\k=\pm1$, it follows that $\Del=0$. Then two of the eigenvalues in \Eq{eigenv} coalesce into $-\g$ to give two third-order exceptional points in this subspace. 
%{\color{blue}
The corresponding eigenvectors are listed in App.~\ref{AppEigev}. There, it is also shown that the three eigenvectors indeed coalesce at the exceptional points.
%}

The solutions exhibit similar behaviors to a damped oscillator discussed in App.~\ref{AppDampOsc}.
Using an analogy to the damped oscillator described there, the evolution of the qubits is critically damped at the exceptional points.
As $|\k|$ increases from below 1, the motion develops from oscillatory (underdamped) to purely exponential decay (overdamped), after crossing over the exceptional point at $|\k|=1$.

%{\color{blue}
The two regions separated by the exceptional points are related to parity-time ($PT$) symmetry phase and spontaneously broken symmetry phase \cite{Bender02,Ozdemir19,ZhangXinchen24}.
$PT$ symmetry could arise when there is a balance between the gain and loss components in non-Hermitian systems.
In the present model, it is convenient to consider $PT$ symmetry in the subspace of the correlation functions $\<\s_\pm\m_\pm\>$ of the two qubits. We present the details in App.~\ref{AppPT}.
By absorbing an overall exponential factor into the correlation functions \cite{Ozdemir19}, we can bring the model with two decay channels into a form that is balanced in gain and loss (see the $\pm i\k$ terms along the diagonal of the time evolution matrix \eqref{M}); hence, the term ``passive" $PT$ symmetry.
The $|\k|<1$ region that exhibits oscillatory motion is in the $PT$-symmetry phase, whereas the $|\k|>1$ region that shows purely exponential decay is in the spontaneously broken $PT$-symmetry phase.
The exceptional points are the critical points of the phase transitions.
In the symmetry broken phase, the time evolution operator remains $PT$ symmetric. However, the solutions no longer satisfy the symmetry \cite{Bender02}, see \Eq{PTlam} and the discussion in App.~\ref{AppPT} for the details.
%}

In this work, our main focus is the entanglement generation dynamics at the initial stage of the time evolution.
We use concurrence $C$ \cite{Wootters98} as a measure of entanglement between the qubits. In this model, it can be determined by $C=2\max\big(0,\sqrt{\lam_1}-\sqrt{\lam_2}-\sqrt{\lam_3}-\sqrt{\lam_4}\big)$, where $\lam_i'$s are the eigenvalues of $\rho (\s_y\m_y)\rho^*\cdot(\s_y\m_y)$ arranged in descending order. 
For the $X$ state \eqref{Xst}, the concurrence has a simple expression \cite{Yu10},
%%%
\begin{align}   \label{C}
    C&=2\max\big(0,|m|-\sqrt{ad},|h|-\sqrt{bc}\big) \,.
\end{align}
%%%

%%%%%%%%%%%%%%%%%%%%%%%%%%
%       Section          %
%%%%%%%%%%%%%%%%%%%%%%%%%%
\section{Initial condition with singly excited qubit}
\label{SecIni10}

For a non-entangled initial condition $|10\>$, the density matrix $\rho_{10}\equiv|10\>\<10|$ begins with $b_0=1$, while the rest of the parameters are zero, $a_0=c_0=d_0=h_0=m_0=m^*_0=0$, or equivalently, $x_0=1=y_0$ and $z_0=0$. 
%{\color{blue}
We define the concurrence as $C_{10}$, consistent with the notation $\rho_{10}$ for the first qubit excited state. 
%}
In this situation, because $a$ and $h$ always vanish, the concurrence depends only on the magnitude of $z$, 
%%%
\begin{align}   \label{C10}
    C_{10}(\t)&=|z|\,.
\end{align}
%%%
For $1<|\k|$, $z$ \eqref{solz} can be written as,
%%%
\begin{align}   \label{z10}
    z&=e^{-\g \t} \frac{i}{\Del^2}
        \bigg(\k
        \cosh(\Del\t)-\k+\Del\sinh(\Del\t)
        \bigg)\,.
\end{align}
%%%
A corresponding expression exists for $-1<\k<1$ when we introduce a parameter $\bD\equiv\sqrt{1-\k^2}$ \eqref{barDel} defined from $\Del=i\bD$, and make use of the identities of hyperbolic functions with imaginary arguments to get functions with oscillatory behaviors.

%%%
\begin{figure}[t]
\centering
\includegraphics[width=3.2in, trim = 4.4cm 10.4cm 3.6cm 10cm]{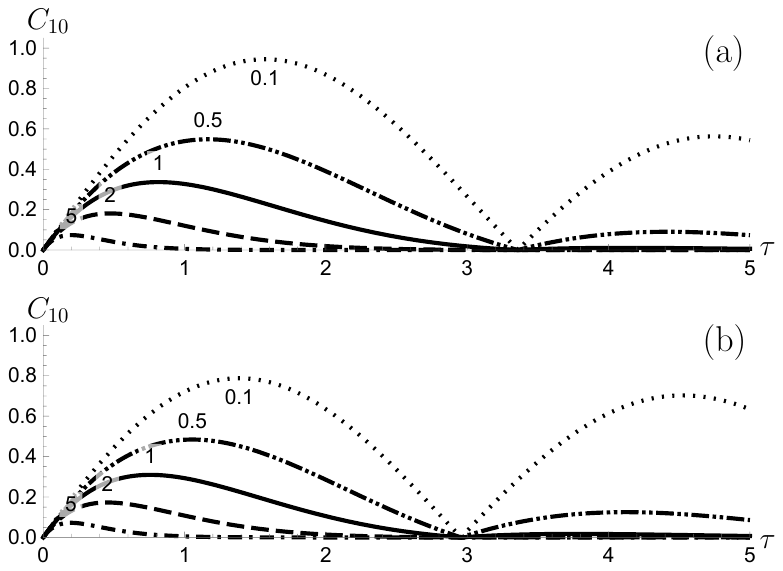}
\caption{Time evolution of concurrence for (a) $\k=0.1$ and (b) $\k=-0.1$ with initial condition $|10\>$. The values of $\g$ are shown on the curves.}
\label{fig1}
\end{figure}
%%%
%%%
\begin{figure}[t]
\centering
\includegraphics[width=3.2in, trim = 4.4cm 10.4cm 3.6cm 10cm]{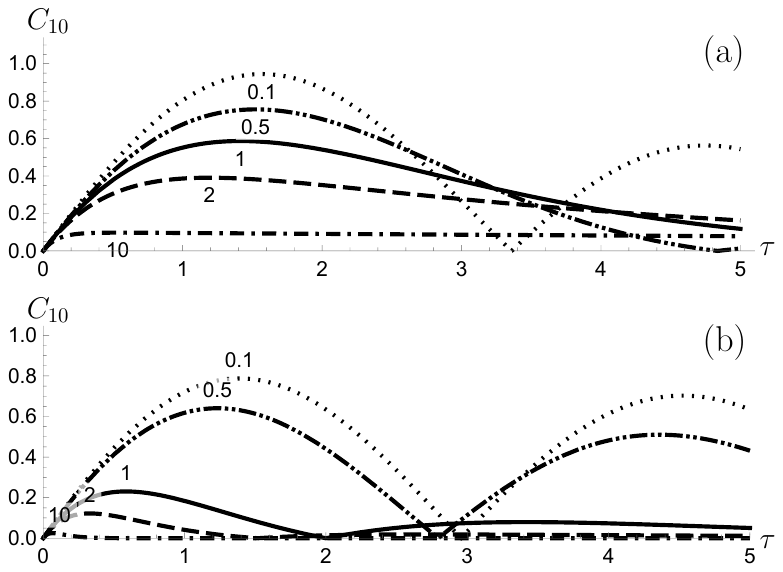}
\caption{Time evolution of concurrence when we set $\g=|\k|$ with initial condition $|10\>$. The values of $\g$ are shown on the curves. (a) $\k>0$ so that $\g_1$ is always 0 and $\g=\g_2$. (b) $\k<0$ so that $\g_2$ is always 0 and $\g=\g_1$. The system is critically damped at $|\k|=1$.}
\label{fig2}
\end{figure}
%%%

For fixed $\k$, \Fig{fig1} (with $\k=0.1$) shows that maximum concurrence occurs when $\g$ assumes its smallest value at $\g=\k$, in view of the constraint $|\k|\leq\g$ \eqref{g<k}.
As $\g$ increases from 0.1 to infinity, the maximum concurrence reduces steadily and approaches 0.
The monotonic behavior can be understood as follows. Since the first qubit does not decay ($\g_1=0$) in the $\g=\k$ situation, there always remains a larger population of the state $|10\>$ in the $\g=\k$ compared to the $\g>\k$ situation. Therefore, there is a higher probability for the state $|10\>$ to swap into $|01\>$ through the coupling between the qubits. Quantum correlation can then be established between the states to generate the maximally entangled Bell states $|10\>\pm|01\>$. 
This explains the maximum value of concurrence for the $\g=\k$ curve.

When total decay rate increases, entanglement generation of Bell's states needs to compete with the higher decay rates of $|10\>$ and $|01\>$ to the ground state $|00\>$.
This results in a lower value of the maximum concurrence generated for the $\g>\k$ situation.
This is revealed in \Fig{fig1}(a) where the $\g=\k$ curve envelops all other $\g>\k$ curves.

Similar behaviors are observed when we fix $\k$ at a negative value, depicted in \Fig{fig1}(b) for $\k=-0.1$. 
The main difference between \Figs{fig1}(a) and \ref{fig1}(b) is that the maximum concurrence achieved with negative $\k$ is less than that with positive $\k$.
This is because for $\g=-\k$, we now have $\g_2=0$. 
A further increase in the total decay rate $\g=\g_1$ will reduce the initial pool of $|10\>$ population, thus slowing down the generation of Bell states.
Hence, the maximum concurrence achieved in the negative-$\k$ curves are less than the positive-$\k$ curves.
Apart from this, \Fig{fig1}(b) shows that the $\g=-\k$ curve envelops all other $\g>-\k$ curves.
We also notice that the revival time of concurrence for negative-$\k$ is shorter than that for positive-$\k$.
In both situations concurrence approaches 1 as $\g\rightarrow0$ when the system returns to unitary.

In \Figs{fig2}(a) and \ref{fig2}(b) we plot the evolution for $\g=\k$ and $\g=-\k$, respectively, for a few $\g'$s.
As $\g$ varies from 0.1 to 10, the system crosses over the exceptional points located at $\k=\pm1$. 
We observe the occurrence of similar differences between \Figs{fig2}(a) and \ref{fig2}(b) with those between \Figs{fig1}(a) and \ref{fig1}(b), that is, (i) monotonic reduction in the maximum concurrence for larger $\g$, (ii) smaller maximum concurrence achieved in negative $\k$ curves compared to positive $\k$ curves, (iii) small revival time for concurrence in negative $\k$ curves, and (iv) the approach of concurrence to 1 as $\g$ vanishes when the system becomes unitary.

If we initiate a setup with the second qubit excited $|01\>$, the concurrence can be deduced from $C_{10}$ \eqref{C10}-\eqref{z10} with a reflection in the parameter $\k\leftrightarrow-\k$, while $\g$ remains the same.
The consequence is a switching of the results between \Figs{fig1}(a) and \ref{fig1}(b), i.e., now \Fig{fig1}(a) applies to negative $\k$ curves, whereas \Fig{fig1}(b) applies to positive $\k$ curves.
Therefore, it is sufficient that we focus on the initial condition $|10\>$ to understand the dynamics.
Furthermore, since the $\g=|\k|$ curve envelops all other $\g>|\k|$ curves for fixed $\k$, in the following we will only consider the $\g=|\k|$ situation when analyzing the maximum concurrence generated across the exceptional points.

%%%
\begin{figure}[t]
\centering
\includegraphics[width=3.2in, trim = 4.4cm 11.cm 4.8cm 10.8cm]{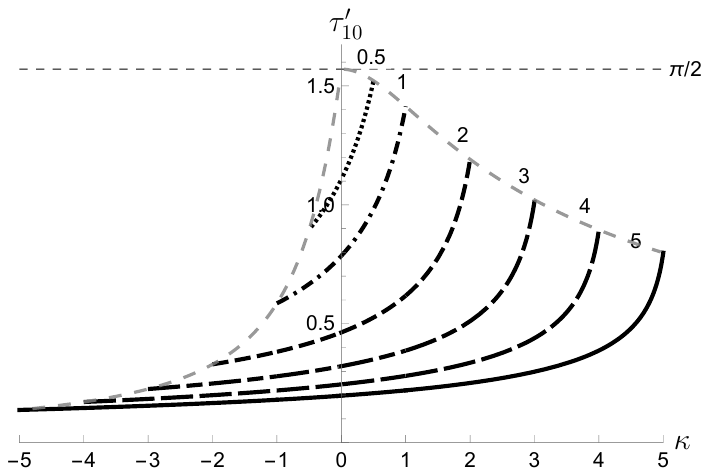}
\caption{Initial condition $|10\>$. The instant ($\t'_{10}$) when the first maximum concurrence for fixed $\g$ occurs. The value of $\g$ ranges from 0.5 to 5, as shown at the end of the curves. $\k$ satisfies the constraint $|\k|\leq\g$ \eqref{g<k}.
Dashed gray lines are the enveloping curves satisfying the condition $\g=|\k|$.}
\label{fig3}
\end{figure}
%%%

Next, we investigate the instant $t'$ at which the system first reaches maximum concurrence. 
It corresponds to the rescaled time $\t'=2Jt'$.
It is the time when the first peaks in \Figs{fig1}(a), \ref{fig1}(b), \ref{fig2}(a) and \ref{fig2}(b) occur.
For the initial condition $|10\>$, we label the instant as $\t'_{10}$. 
The time derivative of the concurrence  $dC_{10}/d\t$ is given in App.~\ref{App1stC10}, see \Eq{dC10dt}. 
Solving $dC_{10}/d\t|_{\t=\t'_{10}}=0$ for the first concurrence maximum, we obtain two expressions for $1<|\k|$ \eqref{t*10<1} and $-1<\k<1$ \eqref{t*10>0} for the overdamped and underdamped region, respectively.
The results are plotted in \Fig{fig3}. 
The curves show the variation of $\t'_{10}$ with $\k$ for fixed value of $\g$ (indicated on the curves).
The longest duration occurs at the right end of each curve when $\k=\g$, a result that can be deduced by inspecting \Fig{fig2}(a).
It is shown at the end of App.~\ref{App1stC10} that the overall longest time to achieve maximum concurrence is $\t'_{10,\text{max}}=\pi/2$, which occurs in the limit $\g\rightarrow 0$, indicated by the horizontal dashed-line in \Fig{fig3}.
This implies that for the initial condition $|10\>$, in all the curves maximum concurrence is reached before
%%%
\begin{align}   \label{t*max}
t'_{10,\text{max}}=\frac{\pi}{4J}\,.
\end{align}
%%%

In \Fig{fig3}, the enveloping dashed grey line on the right (left) quadrant that connects the right (left) end of the curves gives the instant of the first maximum concurrence at $\g=\k$ ($\g=-\k$), respectively. 
A steady and monotonic reduction of the time with the increase of $|\k|$ is evident in the figure.
These curves are the counterparts to a similar curve plotted in \Fig{fig5}(b) for a different initial condition, to be considered in the next section.

%%%%%%%%%%%%%%%%%%%%%%%%%%
%       Section          %
%%%%%%%%%%%%%%%%%%%%%%%%%%
\section{Initial condition with two excited qubits}
\label{SecIni11}

%%%
\begin{figure}[t]
\centering
\includegraphics[width=3.2in, trim = 4.6cm 10.8cm 4.2cm 10.6cm]{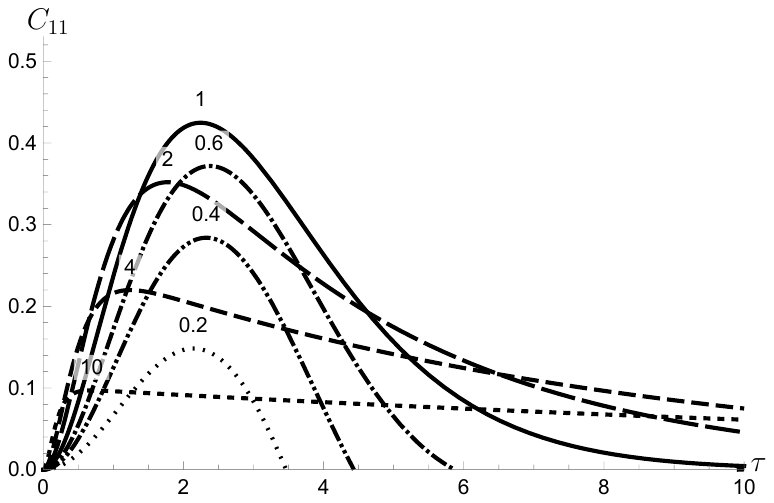}
\caption{$C_{11}(\t)$ with $\k=\g$. The values of $\g$ are shown on the curves. Beginning with $\g=0.2$, at first the increase of dissipation enhances entanglement generation. After reaching a maximum on the curve $\g=1.02$ (almost overlap with the curve $\g=1$, not shown in the figure), a further increase in dissipation starts hindering entanglement generation.}
\label{fig4}
\end{figure}
%%%

For the initial condition with both qubits excited $|11\>$, the initial values of the density matrix $\rho_{11}\equiv|11\>\<11|$ are $a_0=1$, while $b_0=c_0=d_0=h_0=m_0=m^*_0=0$. Since $h$ always vanishes, the concurrence is 
%%%
\begin{align}   \label{C11}
C_{11}(\t)=\max(0,|z|-2\sqrt{ad})\,,
\end{align}
%%%
%{\color{blue}
where we define the concurrence in accordance to the indices of the initial state $\rho_{11}$.
%}
From \Eqs{sola}, \eqref{solz} and \eqref{sold}, we then obtain for $1<|\k|$,
%%%
\begin{align}   \label{C11>1}
&C_{11}(\t)=2e^{-\g\t}\bigg[
\frac{2|\k|}{\k^2-1} \sinh^2\left(\frac{\Del}{2}\t\right) \no
&\quad-\sqrt{\big(1-e^{-\g\t}\big)^2 -\frac{4\k^2 e^{-\g\t}}{\k^2-1}
\sinh^2\left(\frac{\Del}{2}\t\right)}
\bigg]\,.
\end{align}
%%%
Notice that the concurrence is reflection symmetric with respect to $\k\leftrightarrow-\k$.
Therefore, we will consider only positive $\k$ in the following discussion. 
A corresponding expression can be obtained for $-1<\k<1$ in a similar way to $C_{10}$ already discussed at the end of the first paragraph of the last section.

%%%
\begin{figure}[t]
\centering
\includegraphics[width=3.2in, trim = 4.8cm 8.5cm 4cm 11.4cm]{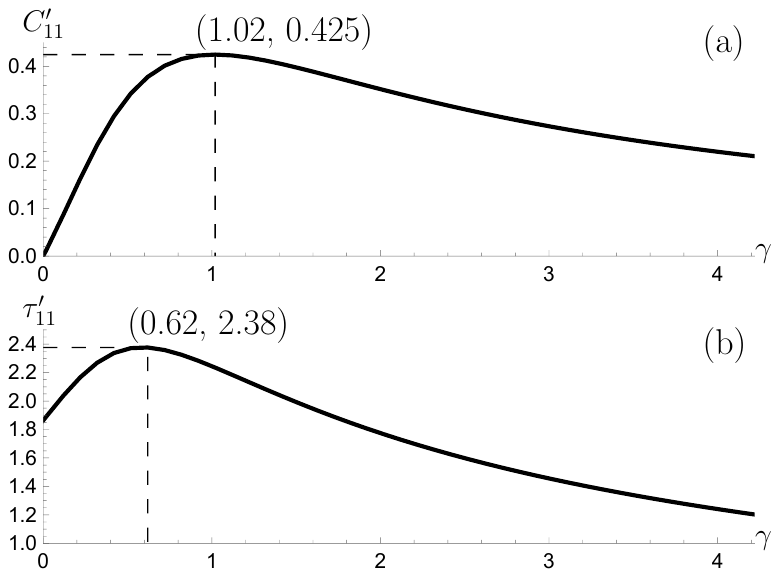}
\caption{(a) The maximum value of $C_{11}(\t)$, $C'_{11}$, is plotted as a function of $\g(=\k)$. The coordinates of the extremum are shown, which are very close to the exceptional point, $\k=1$. (b) The time when the first concurrence maximum is reached, $\t'_{11}$, is plotted as a function of $\g(=\k)$.}
\label{fig5}
\end{figure}
%%%

Similar to \Fig{fig1} in the previous section, the $\g=\k$ curve envelops all other $\k<\g$ curves within it.
Hence, it is sufficient for our purpose to study the time evolution of $\g=\k$, as depicted in \Fig{fig4}.
The curves for a few $\g$ on both sides of the exceptional point are plotted.
%{\color{blue}
Note that for the underdamped cases $\g<1$, there is entanglement sudden death \cite{Yu04,Yu10}, i.e., the concurrence vanishes abruptly in finite time. Because of the definition of concurrence \eqref{C11} that chooses the greater value between 0 and $|z|-2\sqrt{ad}$, the oscillatory behaviors in the underdamped motion are absent from \Fig{fig4} because they occur in the $|z|-2\sqrt{ad}<0$ region.
%}

The first maximum of concurrence achieved in each curve, $C'_{11}$, is plotted in \Fig{fig5}(a) as a function of $\g$, whereas the instant when the first maximum occurs, $\tau'_{11}$, is plotted in \Fig{fig5}(b). 
We observe an interesting behavior in the change of maximum concurrence in \Fig{fig5}(a). 
When the total decay rate increases from 0, the concurrence first increases to an extremum value of $C'_{11}=0.425$ around $\g=1.02$, after it crosses over the exceptional point at $\g=\k=1$ into the overdamped region.
Here in the underdamped region, the increase in dissipation enhances entanglement generation.
The longest duration taken by this process is $\t'_{11}=2.38$, see \Fig{fig5}(b).
Hereafter, a further increase in $\g$ turns the dynamics from enhancing to hindering entanglement generation.
In other words, there is dissipation-assisted entanglement generation for the two-excited-qubits initial condition.
Similar phenomena were found in quantum networks and light-harvesting complexes \cite{Plenio08,Plenio10,Plenio12} where under right condition noise could be assisting energy transport, though the process has a different physical origin from the one considered here.

We can understand how dissipation could assist in entanglement generation as follows. The condition $\g=\k$ implies that the first qubit does not decay $\g_1=0$. 
Hence, decay is solely due to the second qubit, $\g=\g_2$.
When we start with a small decay rate, the initial condition with both qubits excited $|11\>$ does not give much room for the formation of Bell's states, because of the relatively low population of the newly formed $|10\>$ state, which should first swap into $|01\>$ before Bell's states could be formed.
As the dissipation rate gradually increases, the decay of the second qubits produces higher enough population of $|10\>$.
This creates higher probability for the swapping of $|10\>$ into $|01\>$, and thus enhancing entanglement generation between the qubits.
This is an example of dissipation-assisted entanglement conditioned by the initial state.
Since this is a two-step process, i.e., first decay and then swapping, it is not surprising that for a wide range of $\g$, the time taken to generate maximum concurrence in \Fig{fig5}(b) is generally longer than the corresponding time required for the one excited qubit initial condition in \Fig{fig3}, where in the latter it is always less than $\pi/2$.
With the same reasoning, \Fig{fig5}(a) shows that the maximum concurrence in this case is also considerably less than the case achieved with initial condition $|10\>$ in \Figs{fig2}(a) and \ref{fig2}(b) because of the higher loss in populations caused by the longer duration required to generate entanglement.

The maximum concurrence increases steadily until $\k$ is close to the exceptional points where entanglement generation becomes saturated. 
After crossing over the exceptional points into the overdamped region, further increase in the total decay rate ($\g=\g_2$) starts to reduce the population of $|01\>$ needed to form entangled states after the swapping.
Now dissipation begins to hinder entanglement, as commonly occurs in most systems. Thereafter, the maximum concurrence decreases gradually as $\g$ increases further.

%{\color{blue}
Quantum entanglement, quantified by concurrence considered in this work, is a form of quantum correlation. One might wonder whether the extremum behavior shown in \Figs{fig5}(a)-(b)  would also show up in the correlation functions of two qubits $\<\s_i\m_j\>\equiv\tr(\s_i\m_j\rho)$, where $i,j=x,y,z$. Using $\<\s_x\m_y\>=-i(z-h+h^*)$ \eqref{corrxy} as an example, in App.~\ref{AppCorFunc} we show that the extremum behavior indeed occurs in the two excited qubits initial condition, see \Fig{fig7}(a). 
However, since the concurrence $C_{11}$ \eqref{C11} is also affected by the diagonal components of the density matrix, i.e., $a$ and $d$, the extrema in the two quantities do not occur at the same value of $\g$. 
Furthermore, the instant when the first maximum of the correlation function is achieved is monotonously reducing with $\g$, see \Fig{fig7}(b). It behaves differently from the corresponding time in concurrence, see \Fig{fig5}(b). 
Therefore, the correlation functions cannot be used as substitutes of concurrence to identify the maximum of entanglement. 
%}

%%%%%%%%%%%%%%%%%%%%%%%%%%
%       Section          %
%%%%%%%%%%%%%%%%%%%%%%%%%%
\section{Mixture of states}
\label{SecIniMixed}

%%%
\begin{figure}[t]
\centering
\includegraphics[width=3.2in, trim = 4.8cm 10cm 4cm 9.5cm]{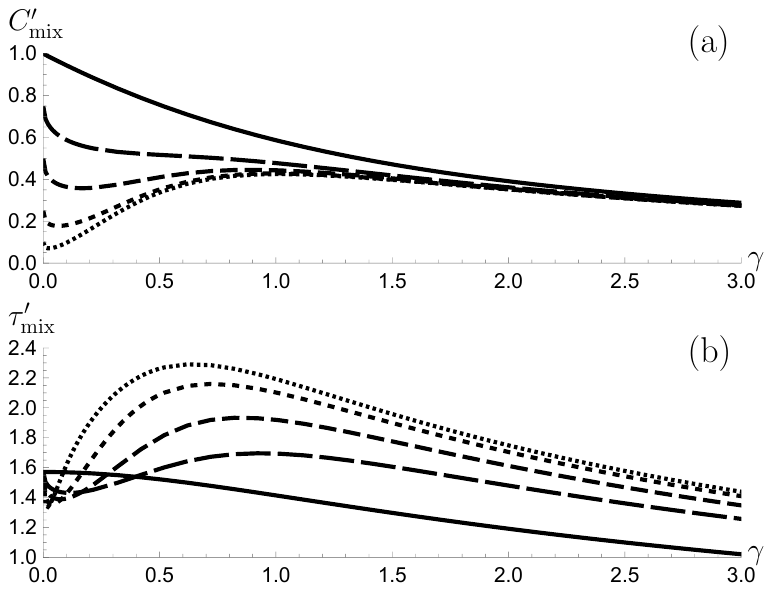}
\caption{Mixed state $\rho_\text{mix}(\al)$. (a) Maximum concurrence is plotted as a function of $\g(=\k)$. (b) The time when the first maximum concurrence occurs. The values of $\al$ are, 1 (solid), 0.75 (long-dashed), 0.5 (medium-dashed), 0.25 (short-dashed), and 0.1 (dotted) curves, respectively. }
\label{fig6}
\end{figure}
%%%

%%%
\begin{figure}[b]
\centering
\includegraphics[width=3.2in, trim = 4.8cm 6.cm 4cm 13.8cm]{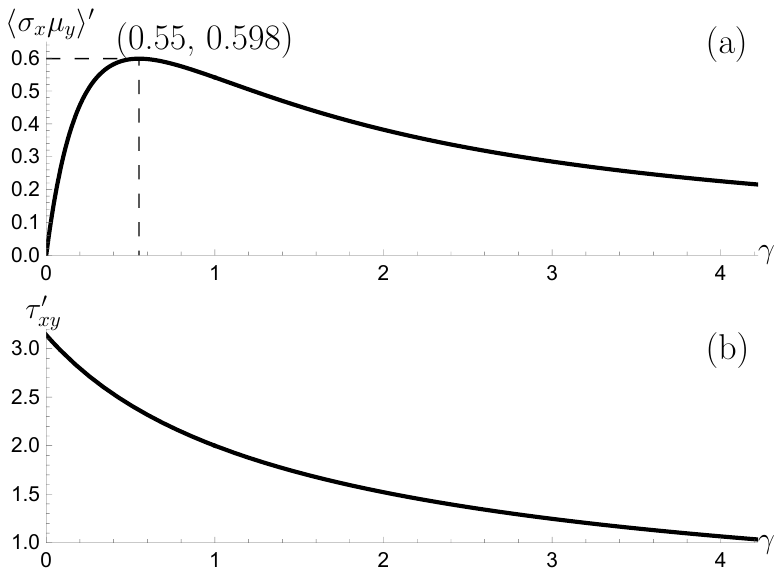}
\caption{(a) The maximum value of the correlation function, $\<\s_x\m_y\>'$, is plotted as a function of $\g(=\k)$. The extremum occurs far from the exceptional point, $\k=1$. (b) The instant when the first maximum of $\<\s_x\m_y\>'$ is reached, $\t'_{xy}$, is plotted as a function of $\g(=\k)$.}
\label{fig7}
\end{figure}
%%%

We can form a mixture of states by a convex sum of the initial conditions of the two qubits, which we define as
%%%
\begin{align}   \label{rhomix}
    \rho_\text{mix}(\al)&\equiv\al\rho_{10}+(1-\al)\rho_{11}\,,
\end{align}
%%%
where $0\leq\al\leq1$. This state has the initial condition $a_0=1-\al, b_0=x_0=y_0=\al$, while the rests are zero.
The concurrence has the same expression as \Eq{C11}, though with different parameters listed in App.~\ref{AppMixedSt}, see \Eqs{amix}-\eqref{dmix}.

The values of the first maximum concurrence in this case, which we denote by $C'_\text{mix}$, which can be achieved for various $\al$ and the time taken, $\t'_\text{mix}$, are plotted in \Figs{fig6}(a) and \ref{fig6}(b), respectively, as a function of the total decay rate $\g$ for the condition $\k=\g$.
Here we only consider positive $\k$ for simplicity to show the general features of entanglement dynamics with $\rho_\text{mix}(\al)$ as the initial condition. 

By comparing \Figs{fig6}(a) and \ref{fig6}(b) with \Figs{fig5}(a) and \ref{fig5}(b) ($\al=0$), respectively, we find a competition between the two entanglement dynamics discussed in the previous sections.
Focusing on \Fig{fig6}(a), we begin with the state $\rho_{10}$ (solid curve with $\al=1$).
As $\al$ reduces from 1, we first observe that the maximum concurrence reduces steadily over a wide range of $\g$ until the weight of the state $\rho_{11}$ becomes prominent enough to bring noticeable changes to the dynamics. This occurs approximately below $\al<0.75$ (long-dashed curve).
Now in the underdamped region $\g=\k<1$ a hollow starts to form which becomes obvious around $\al=0.5$ (medium-dashed curve), signifying a shift from a mode of entanglement hindrance to a mode of entanglement enhancement by dissipation.
Now, maxima in the concurrence curves start to appear, and there exists optimal dissipation rate for entanglement generation over a wide range of mixed states.
\Fig{fig6}(b) shows that the time needed to achieve maximum concurrence generally increases as the weight of $\rho_{11}$ increases, consistent with the discussion in the previous section.
Here, optimal entanglement generation is compromised by the need of a longer interval to achieve its maximum.

%%%%%%%%%%%%%%%%%%%%%%%%%%
%       Section          %
%%%%%%%%%%%%%%%%%%%%%%%%%%
\section{Conclusion}
\label{SecConc}

We have shown that the interplay between the configuration of the initial states and the interaction between the two qubits is crucial in deciding the character of entanglement dynamics at the initial stage of evolution in a two-qubit open quantum system.
%{\color{blue}
Contrary to the common belief that dissipation is in general detrimental towards entanglement generation, for certain initial states an increase in the total decay rate can instead favor entanglement generation.
%}

We note that the configuration that shows the crossover behavior requires one of the excitations to first relax before entanglement can be generated.
This causes the system to undergo dissipation throughout a relatively longer time scale of the order of the qubit's decay time.
As a result, the overall maximum entanglement achieved is relatively less.
In summary, we have clarified the general features required and the regions the system must lie in to optimize entanglement generation across the exceptional points in the two-qubit open quantum system.
%{\color{blue}
Since dissipation is inevitable in quantum processes, our results provide a different perspective on the effects of dissipation to entanglement dynamics.
%}

\acknowledgments
We are grateful for the support of the Fundamental Research Grant Scheme (FRGS), Grant No.~FRGS/1/2020/STG07/UNIM/02/01, by the Ministry of Higher Education Malaysia (MOHE). Y.S.H. is supported as a Graduate Research Assistant (GRA) under this grant.

\appendix

%%%%%%%%%%%%%%%%%%%%%%%%%%
%       Section          %
%%%%%%%%%%%%%%%%%%%%%%%%%%
\section{Classical damped oscillator}
\label{AppDampOsc}

The position of a damped oscillator obeys the equation
$m\ddot{x}+\g \dot{x}+kx=0$, where $\g$ and $k$ denotes the damping and the spring constant, respectively. We can turn the equation into two simultaneous equations. In matrix form, they are
%%%
\begin{align}   \label{classosceq}
   \dot{\boldsymbol{X}}
            &=
    \boldsymbol{K}\cdot \boldsymbol{X}
   \,.
\end{align}
%%%
where
%%%
\begin{align}   \label{X}
    \boldsymbol{X}\equiv \left(\begin{array}{c}
            x \\
            v 
            \end{array}\right)
\end{align}
%%%
and
%%%
\begin{align}   \label{k}
    \boldsymbol{K}&=\left(\begin{array}{cc}
            0 &1 \\
            -k/m &-\g/m
            \end{array}\right)
\end{align}
%%%
with the solution $\boldsymbol{X}(t)=e^{\boldsymbol{K}t}\cdot\boldsymbol{X}$. The eigenvalue equation $\boldsymbol{K}\cdot\boldsymbol{v}=\tilde{\lam}\boldsymbol{v}$ has the solutions (we set $m=1$)
%%%
\begin{align}   \label{eigK}
    \tilde{\lam}_\pm&=-\frac{1}{2}\big(\g\pm\tilde{\Del}\big)\,,
\end{align}
%%%
where
%%%
\begin{align}   \label{Delt}
    \tilde{\Del}&\equiv\sqrt{\g^2-4k}\,,
\end{align}
%%%
and the eigenvectors
%%%
\begin{align}   \label{eigvK}
    \boldsymbol{v}_\pm&=    
        \left(\begin{array}{c}
            \tilde{\lam}_\mp \\
            -k
            \end{array}\right)\,.
\end{align}
%%%

The damped oscillator can show three types of motion. The motion is oscillatory when $\g^2<4k$, where $\tilde{\Del}$ is imaginary, i.e., the oscillator is underdamped.
The position $x$ decays exponentially when $\g^2>4k$, where $\tilde{\Del}$ is real. The oscillator is then overdamped.
Critical damping occurs at $\g^2=4k$, when $\tilde{\Del}=0$.

The eigenvalues and eigenvectors coalesce at the exceptional point, which occurs at $\g^2=4k$. This corresponds to critical damping.
The eigenvalues coalesce to 
%%%
\begin{align}   \label{lamc}
    \tilde{\lam}_\text{c}\equiv\tilde{\lam}_+=\tilde{\lam}_-=-\frac{\g}{2}\,.
\end{align}
%%%
At the exceptional point, the first eigenvector
%%%
\begin{align}   \label{EPeigev}
    \boldsymbol{v}_0=        
        \left(\begin{array}{c}
            1 \\
            -\tilde{\lam}_\text{c}
            \end{array}\right)
\end{align}
%%%
satisfies the eigenvalue equation
%%%
\begin{align}   \label{EPeigEq}
    \boldsymbol{K}\cdot\boldsymbol{v}_0&=\tilde{\lam}_\text{c}\boldsymbol{v}_0\,.
\end{align}
%%%
To complement $\boldsymbol{v}_0$, we need to consider the generalized eigenvalue equation \eqref{EPeigEq},
%%%
\begin{align}   \label{geneigev}
        \boldsymbol{K}\cdot\boldsymbol{v}_1&=\tilde{\lam}_\text{c}\boldsymbol{v}_1
        + \boldsymbol{v}_0\,.
\end{align}
%%%
$\boldsymbol{v}_0$ and $\boldsymbol{v}_1$ are the generalized eigenvectors of $\boldsymbol{K}$ at the exceptional point.
There is a great deal of freedom in the choice of the second generalized eigenvector
%%%
\begin{align}   \label{v1}
    \boldsymbol{v}_1&=        
            \left(\begin{array}{c}
            a \\
            -1-a\tilde{\lam}_\text{c}
            \end{array}\right)\,,
\end{align}
%%% 
for real $a$.

%%%%%%%%%%%%%%%%%%%%%%%%%%
%       Section          %
%%%%%%%%%%%%%%%%%%%%%%%%%%
\section{Solutions to equations of motion}
\label{AppSol}

The equations of motion of the matrix elements in the $X$ state \eqref{Xst} are
%%%
%\begin{subequations}
\begin{align}   \label{at}
        \dot{a}&=-2\g a\,,
\end{align}
%%%
%%%
\begin{align}   
    \dot{x}&=2\g a-\g x+\k y\,,
%        \dot{b}+\dot{c}&=2\g a-\g(b+c)+\k(b-c)\,,\\
\end{align}
%%%
%%%
\begin{align}   
    \dot{y}&=2\k a+\k x-\g y+iz\,,
%        \dot{b}-\dot{c}&=2\k a+\k(b+c)-\g(b-c)+i(m-m^*)\,,\\
\end{align}
%%%
%%%
\begin{align}   \label{zt}
        \dot{z}&=-\g z+i y\,, 
%        \dot{m}-\dot{m}^*&=-\g(m-m^*)+i(b-c)\,,\\
\end{align}
%%%
%%%
\begin{align}   
        \dot{d}&=\g x-\k y\,,
\end{align}
%%%
%%%
\begin{align}   \label{mmt}
        \dot{m}+\dot{m}^*&=-\g(m+m^*)\,,
\end{align}
%%%
%%%
\begin{align}   \label{ht}
        \dot{h}&=-\g h\,,
\end{align}
%\end{subequations}
%%%
where dots label time derivative with respect to $\t$.
We have omitted the time dependence on the parameters for simplicity.
In the equation, $\g$ and $\k$ are the total decay rate and difference in the decay rates between the qubits, defined in \Eqs{gam} and \eqref{kap}, respectively.

The set of equations has the following solutions, where $x,y$ and $z$ are defined in \Eqs{x}--\eqref{z}, respectively.
In the following expressions, the subscript 0 denotes initial conditions, 
%%%
\begin{align}   \label{sola}
    a(\t)&=e^{-2\g \t}a_0\,,
\end{align}
%%%
%%%
\begin{align}   \label{solmm}
    m(\t)+m^*(\t)&=e^{-\g \t}(m_0+m_0^*)\,,
\end{align}
%%%
%%%
\begin{align}   \label{solh}
    h(\t)&=e^{-\g \t}h_0\,,
\end{align}
%%%
%%%
\begin{widetext}
%%%
\begin{align}   \label{solx}
    x(\t)&=-2a_0 \left[ e^{-2\g\t}-e^{-\g\t} \left(1-\frac{\k^2}{\Del^2} \big(\cosh(\Del\t)-1\big)\right) \right]
        +x_0 e^{-\g\t} \left(1-\frac{\k^2}{\Del^2} \big(\cosh(\Del\t)-1\big)\right)   \no
        &\quad   
    +y_0\frac{\k}{\Del} e^{-\g\t} \sinh(\Del\t)
    +z_0 \frac{i\k}{\Del^2} e^{-\g\t} 
    \big(\cosh(\Del\t)-1\big)  \,,
\end{align}
%%%
%%%
\begin{align}   \label{soly}
    y(\t)&=y_0 e^{-\g\t} \cosh(\Del\t)
    +\left(a_0\frac{2\k}{\Del} +x_0\frac{\k}{\Del}
    +z_0 \frac{i}{\Del}\right) e^{-\g\t} 
    \sinh(\Del\t)  \,,
\end{align}
%%%
%%%
\begin{align}   \label{solz}
    z(\t)&=(2a_0+x_0) \frac{i\k}{\Del^2} e^{-\g\t} 
    \big(\cosh(\Del\t)-1\big) 
    +y_0 \frac{i}{\Del} e^{-\g\t} \sinh(\Del\t)
    +z_0 e^{-\g\t} \left(1-\frac{1}{\Del^2}
            \big(\cosh(\Del\t)-1\big)  \right)\,,
\end{align}
%%%
%%%
\begin{align}   \label{sold}
    d(\t)&=d_0+a_0\left[1+e^{-2\g\t}
        -2e^{-\g\t}  \left(1-\frac{\k^2}{\Del^2} \big(\cosh(\Del\t)-1\big)\right) \right]
        +x_0\left[1-e^{-\g\t} 
        \left(1-\frac{\k^2}{\Del^2} \big(\cosh(\Del\t)-1\big)\right) \right] \no
        &\quad
        -y_0\frac{\k}{\Del} e^{-\g\t} \sinh(\Del\t) 
        -z_0 \frac{i\k}{\Del^2} e^{-\g\t} 
        \big(\cosh(\Del\t)-1\big)  \,.
\end{align}
%%%
\end{widetext}

%%%%%%%%%%%%%%%%%%%%%%%%%%
%       Section          %
%%%%%%%%%%%%%%%%%%%%%%%%%%
\section{%\color{blue} 
Third-order exceptional points}
\label{AppEigev}

In this appendix, we show that the five-parameter subspace represented by the vector $\u=(a,x,y,z,d)^\text{T}$, where $\text{T}$ denotes matrix transpose, has two third-order exceptional points at $\k=\pm1$. The matrix equation $\dot{\u}=\L\cdot\u$ has the non-Hermitian time evolution generator
%%%
\begin{align}   \label{G}
    \L\equiv\left(\begin{array}{ccccc}
               -2\g &0 &0 &0 &0 \\
               2\g &-\g &\k &0 &0 \\
               2\k &\k &-\g &i &0 \\
               0 &0 &i &-\g &0 \\
               0 &\g &-\k &0 &0
             \end{array} \right)\,.
\end{align}
%%%
It has the eigenvalues listed in \Eq{eigenv}, where $\lam_0$ corresponds to the equilibrium state and $\Delta$ is defined in \Eq{Del}. The corresponding eigenvectors are
%%%
\begin{align}   \label{eigv}
    \u_0&=(0,0,0,0,1)^\T\,,\\
    \u_1&=(1,-2,0,0,1)^\T\,,\\
    \u_2&=(0,1,0,i\k,-1)^\T\,,\\
    \u_3&=\pm(0,\k,-\Del,i,-\k)^\T\,,\\
    \u_4&=\pm(0,\k,\Del,i,-\k)^\T\,,
\end{align}
%%%
where in $\u_3$ and $\u_4$, we use the sign $\pm$ when $\k=\pm1$. 
By inspection, as $\k\rightarrow\pm1$ at the exceptional points, $\u_2,\u_3$ and $\u_4$ coalesce into
%%%
\begin{align}
    \u_2=\u_3=\u_4=(0,1,0,\pm i,-1)^\T\,.
\end{align}
%%%
For non-Hermitian systems, we also need to consider the left-eigenvalue problem $\L^\dagger\cdot \boldsymbol{v}=\lam \boldsymbol{v}$. The eigenvectors satisfy the biorthogonal relation $\boldsymbol{v}_i^\dagger\cdot \u_j=\delta_{ij}$ and the completeness relation $\sum_i \boldsymbol{u}_i\cdot\boldsymbol{v}^\dagger_i=I$, where $i,j$ range from 0 to 4. We will not list the left eigenvectors here since this is not along the main theme of this work.

%%%%%%%%%%%%%%%%%%%%%%%%%%
%       Section          %
%%%%%%%%%%%%%%%%%%%%%%%%%%
\section{%\color{blue} 
$PT$-symmetry and symmetry broken phase}
\label{AppPT}

In this appendix, we discuss the $PT$-symmetry phase transition of the two-qubit model across the exceptional points \cite{Bender02,ZhangXinchen24}. 
To illustrate the phase transition, it is convenient to consider the expectation values of the four operators, $\<\s_\pm\m_\pm\>$, where we define $\<O\>\equiv\tr(O\rho)$, and $O$ is a two-qubit operator. We calculate the time evolution of the operators by tracing out the operators with the GKSL master equation \eqref{rhot}. In terms of the vector
%%%
\begin{align}   \label{c}
   \q\equiv\big( \<\s_+\m_+\>, \<\s_+\m_-\>, \<\s_-\m_+\>, \<\s_-\m_-\> \big)^\T \,,
\end{align}
%%%
where $\T$ denotes matrix transpose, we obtain the equation of motion $d\q/d\t=-\g\q-i\M\cdot\q$, where 
%%%
\begin{align}   \label{M}
  \M\equiv\left(\begin{array}{cccc}
             i\k & \frac{1}{2} & -\frac{1}{2} & 0 \\
             \frac{1}{2} & 0 & 0 & -\frac{1}{2} \\
             -\frac{1}{2} & 0 & 0 & \frac{1}{2} \\
             0 & -\frac{1}{2} & \frac{1}{2} & -i\k 
           \end{array}
            \right)
\end{align}
%%%
is a time evolution matrix.
Redefining the vector as $\bar{\q}\equiv \exp(\g\t)\q$, which is called a gauge transformation in Ref.~\cite{Ozdemir19}, we obtain an explicit $PT$-symmetric equation of motion
%%%
\begin{align}   \label{PTeq}
  i\frac{d}{d\t}\bar{\q}=\M\cdot\bar{\q}\,.
\end{align}
%%%

To show that \Eq{PTeq} is indeed $PT$ symmetric, we define the parity operator as $P\equiv\s_x\m_x$, and the antilinear time-reversal operator $T$ is a complex conjugation operator with the action $TiT=-i$, and reverses the time $T\t T=-\t$. Furthermore, $P$ and $T$ commute $[P,T]=0$ and they are their own inverse, $P^2=I$ and $T^2=I$, where $I$ is the identity operator. We can now verify that $\M$ is indeed $PT$ symmetric
%%%
\begin{align}   \label{PTM}
    PT\M(PT)^{-1}=\M\,,
\end{align}
%%%
and \Eq{PTeq} shows this as well.

The eigenvalue problem $\M\cdot\v=\lam\v$ has four solutions, $\lam_1=0, \lam_2=0, \lam_3=-\bD$, and $\lam_4=\bD$, where
%%%
\begin{align}   \label{barDel}
    \bD&\equiv\sqrt{1-\k^2}\,.
\end{align}
%%%
The corresponding eigenvectors are
%%%
\begin{align}   \label{Mv}
    \v_1&=\left(\begin{array}{c}
                  1 \\
                  -2ik \\
                  0 \\
                  1 
                \end{array}\right),    
    & 
    \v_2&=\left(\begin{array}{c}
                  0 \\
                  1 \\
                  1 \\
                  0 
                \end{array}\right),\\
    \v_3&=\left(\begin{array}{c}
                  1-2\bD^2+2ik\bD \\
                  \bD-i\k \\
                  -(\bD-i\k) \\
                  1 
                \end{array}\right),
    &
     \v_4&=\left(\begin{array}{c}
                  1-2\bD^2-2ik\bD \\
                  -(\bD+i\k) \\
                  \bD+i\k \\
                  1 
                \end{array}\right).
\end{align}
%%%

The eigenvalues are real when $|\k|<1$. This is the $PT$-symmetric region where the eigenvalues are also $PT$-symmetric, i.e., $PT\lam_i(PT)^{-1}=\lam_i$, though $\v_i$ may not be the eigenstates of the $PT$ operator. Apart from an overall exponential decay, the motion has oscillatory behavior, $\q(\t)=\exp(-\g\t)\sum_i c_i \exp(-i\lam_i\t)\v_i$ with real $\lam_i$, where we expand the initial condition  in terms of the eigenvectors $\q(0)=\sum_i c_i\v_i$, and $c_i'$s are the expansion coefficients. This region is the analog of the underdamped motion of the damped oscillator discussed in App.~\ref{AppDampOsc}.

In the region $|k|>1$, two of the eigenvalues $\lam_{3,4}$ become imaginary $\lam_3=-i\Del$ and $\lam_4=i\Del$, where $\Del=\sqrt{\k^2-1}$ \eqref{Del}. In this region, even though $\M$ remain $PT$ symmetric, two of the eigenvalues are not, 
%%%
\begin{align}   \label{PTlam}
PT\lam_{3,4}(PT)^{-1}=\lam^*_{3,4}\neq\lam_{3,4}\,,
\end{align}
%%%
i.e., the $PT$-symmetry of the solutions is spontaneously broken \cite{Bender02,Ozdemir19}. The motion is now purely exponential, in analogy to the overdamped motion of a damped oscillator. The critical points where the transition between the $PT$-symmetry phase and spontaneously broken $PT$-symmetry phase occur are the exceptional points. In this model, they are $\k=\pm1$. In this subspace of four operators, the exceptional points are of second order, in which the pair of eigenvalues $\lam_3$ and $\lam_4$ as well as the pair of eigenvectors $\v_3$ and $\v_4$ coalesce.

%%%%%%%%%%%%%%%%%%%%%%%%%%
%       Section          %
%%%%%%%%%%%%%%%%%%%%%%%%%%
\section{Instant of the first maximum concurrence}
\label{App1stC10}

In this appendix we obtain the instant defined as $\t'_{10}$, when the first maximum concurrence of $C_{10}$ is reached.
The time derivative of the concurrence is
%%%
\begin{align}   \label{dC10dt}
    \frac{dC_{10}}{d\t}&=2\frac{|\k|}{\k^2-1}
    \bigg|\g+ A \cosh(\Del\t) + B \sinh(\Del\t)\bigg|\,,
    \end{align}
%%%
where
%%%
\begin{align}   \label{A}
    A&=\frac{\Del^2}{\k}-\g\,,%\frac{1}{\k}-\k
\end{align}
%%%
%%%
\begin{align}   
    B&=\left(1-\frac{\g}{\k}\right)\Del\,.
\end{align}
%%%
Solving $dC_{10}/d\t|_{\t=\t'_{10}}=0$ for the first maximum, we obtain for $1<|\k|$,
%%%
\begin{align}   \label{t*10<1}
    \tanh \big(\t'_{10}\Del\big)&=\frac{\g|\k|B\mp\Del A\sqrt{1+\g^2}}
    {\g|\k|A\mp \Del B\sqrt{1+\g^2}}\,,
\end{align}
%%%
where we choose the upper sign in \Eq{t*10<1} for $1<\k$, and the lower sign for $\k<-1$.
For $-1<\k<1$, we have instead
%%%
\begin{align}   \label{t*10>0}
    \tan(\t'_{10}\bD)
    &=
   \frac{\g|\k|\bar{B}
   \mp \bD A\sqrt{1+\g^2}}
   {\g|\k|A\pm \bD\bar{B}\sqrt{1+\g^2}}
    \,,
\end{align}
%%%
%%%
\begin{align} 
    \bar{B}&\equiv\left(1-\frac{\g}{\k} \right)\bD\,,
\end{align}
%%%
where $\bD$ is defined in \Eq{barDel} and we choose the upper sign in \Eq{t*10>0} for $-1<\k<0$ and the lower sign for $0<\k<1$.
At the exceptional points, $\k=\pm1$, we have a simple expression for the instant of the first maximum concurrence, given by
%%%
\begin{align}   \label{t*10EP}
    \t'_{10,\pm}&= \frac{1}{\g}\left(
    1\mp \g\pm\sqrt{1+\g^2}\right)\,.
\end{align}
%%%

When the decay rates of both qubits are the same, i.e., $\k=0$, \Eq{t*10>0} goes into $\tan\t'_{10}\rightarrow1/\g$. The maximum concurrence then occurs at $\t'_{10}|_{\k=0}=\tan^{-1}(1/\g)$.
For fixed $\g$, $\t'_{10}$ is largest when $\k=\g$, see the discussion in Sec.~\ref{SecIni10}.
The overall maximum time is then reached in the limit $\g\rightarrow 0$, where
%%%
\begin{align}   \label{tau*max}
    \t'_{10,\text{max}}\rightarrow \frac{\pi}{2}\,,
\end{align}
%%%
or equivalently, $t'_{10,\text{max}}=\pi/4J$ \eqref{t*max}.
The result can be obtained from \Eq{t*10>0} by first setting $\k=\g$. Then, we substitute $\bar{B}=0$ so that \Eq{t*10>0} goes into $\tan(\sqrt{1-\g^2}\t'_{10,\text{max}})=1/\g^2$. In the limit $\g\rightarrow0$, we arrive at \Eq{tau*max}.

%%%%%%%%%%%%%%%%%%%%%%%%%%
%       Section          %
%%%%%%%%%%%%%%%%%%%%%%%%%%
\section{%\color{blue} 
Correlation functions}
\label{AppCorFunc}

In this appendix we show that the extremum behavior of the concurrence, $C'_{11}$, depicted in \Fig{fig5}(a) for the initial condition of two excited qubits also shows up in the two-qubit correlation function $\<\s_x\m_y\>=\tr(\s_x\m_y\rho)$. The quantity is related to the off-diagonal component of the $X$ state by
%%%
\begin{align}   \label{R12}
    \<\s_x\m_y\>=-i(z-h+h^*)\,.
\end{align}
%%%
For the initial condition with two excited qubits $\rho_{11}$, cf.~the first paragraph of Sec.~\ref{SecIni11}, since $h(t)=h^*(t)=0$, it takes the form
%%%
\begin{align} \label{corrxy}
    \<\s_x\m_y\>=-iz=\frac{2\k}{\Del^2}e^{-\g \t}(\cosh\Del\t-1)\,,
\end{align}
%%%
where $\Del=\sqrt{\k^2-1}$ is defined in \Eq{Del}. 
We can solve for the instant when the first maximum is reached, denoted by $\tau'_{xy}$, from $d\<\s_x\m_y\>/d\t|_{\t=\t'_{xy}}=0$, to get
%%%
\begin{align} \label{taucorrxy}
    \coth\left(\frac{\Del}{2} \t'_{xy}\right)=\frac{\g}{\Del}\,.
% \t'&=\left\{\begin{array}{cc}
%              \displaystyle \frac{1}{\bD}\tan^{-1}\frac{2\g\bD}{\g^2-\bD^2}\,,  &\quad |\k|<1\,,\\\\
%              \displaystyle \frac{2}{\g}\,, &\quad \k=\pm1 \,,\\\\
%              \displaystyle \frac{1}{\Del}\ln\frac{\g+\Del}{\g-\Del}\,,  &\quad |\k|>1\,, \\
%             \end{array}\right.
\end{align}
%%%
For the special case $\k=\g$, the first maximum $\<\s_x\m_y\>'$ as a function of $\g$ has an extremum, see \Fig{fig7}(a), which is far from the exceptional point. However, the instant, $\t'_{xy}$, at which the maximum is reached is a monotonically reducing function in $\g$, see \Fig{fig7}(b). 
\Figs{fig7}(a) and \ref{fig7}(b) exhibit slightly different behaviors from \Figs{fig5}(a) and \ref{fig5}(b) because the concurrence $C_{11}$ \eqref{C11} not only depends on the correlation function differently compared to $\<\s_x\m_y\>$ \eqref{R12}, it also depends on the diagonal components of the density matrix.

We note that $\<\s_y\m_x\>=i(z+h-h^*)$. For the initial condition $\rho_{11}$, $\<\s_y\m_x\>=-\<\s_x\m_y\>$. Therefore, the plot of the first extremum of $\<\s_y\m_x\>$, denoted by $\<\s_y\m_x\>'$, is the reflection of \Fig{fig7}(a) along the horizontal $\g$ axis, whereas the instant when the first extremum of $\<\s_y\m_x\>$ occurs, denoted by $\t'_{yx}$, gives the same plot as in \Fig{fig7}(b).

%%%%%%%%%%%%%%%%%%%%%%%%%%
%       Section          %
%%%%%%%%%%%%%%%%%%%%%%%%%%
\section{Mixed states}
\label{AppMixedSt}

For the mixed state $\rho_\text{mix}(\al)$ \eqref{rhomix}, substituting the initial conditions $a_0=1-\al, x_0=y_0=\al$, and $c_0=d_0=z_0=h_0=0$, into \Eqs{sola}, \eqref{solz} and \eqref{sold}, we obtain for $1<|\k|$,
%%%
\begin{align}   \label{amix}
    a(\t)=(1-\al) e^{-2\g \t}\,,
\end{align}
%%%
%%%
\begin{align}   \label{zmix}
    z(\t)=2i e^{-\g \t}\sinh^2\left(\frac{\Del}{2}\t\right)
        \left[(2-\al)\frac{\k}{\Del^2}
        +\frac{\al}{\Del}\coth\left(\frac{\Del}{2}\t\right)\right]\,,
\end{align}
%%%
%%%
\begin{align}   \label{dmix}        
    d(\t)&=(1-\al)\left[ (1-e^{-\g\t})^2
        -\frac{4\k^2}{\Del^2} e^{-\g\t} \sinh^2\left(\frac{\Del}{2}\t\right) \right] \no
        &\quad+\al \bigg[ 1-e^{-\g\t} -2e^{-\g\t} \sinh^2 \left(\frac{\Del}{2}\t\right) \no 
        &\qquad\qquad\qquad
        \times \left( \frac{\k^2}{\Del^2}
        +\frac{\k}{\Del} \coth\left(\frac{\Del}{2}\t\right) \right)\bigg]\,.
\end{align}
%%%
A corresponding expression for $-1<\k<1$ can be obtained using the identities of hyperbolic functions with imaginary arguments.

%\bibliography{2qubitsET}{}
%apsrev4-2.bst 2019-01-14 (MD) hand-edited version of apsrev4-1.bst
%Control: key (0)
%Control: author (8) initials jnrlst
%Control: editor formatted (1) identically to author
%Control: production of article title (0) allowed
%Control: page (0) single
%Control: year (1) truncated
%Control: production of eprint (0) enabled
\providecommand{\noopsort}[1]{}\providecommand{\singleletter}[1]{#1}%

\end{document}